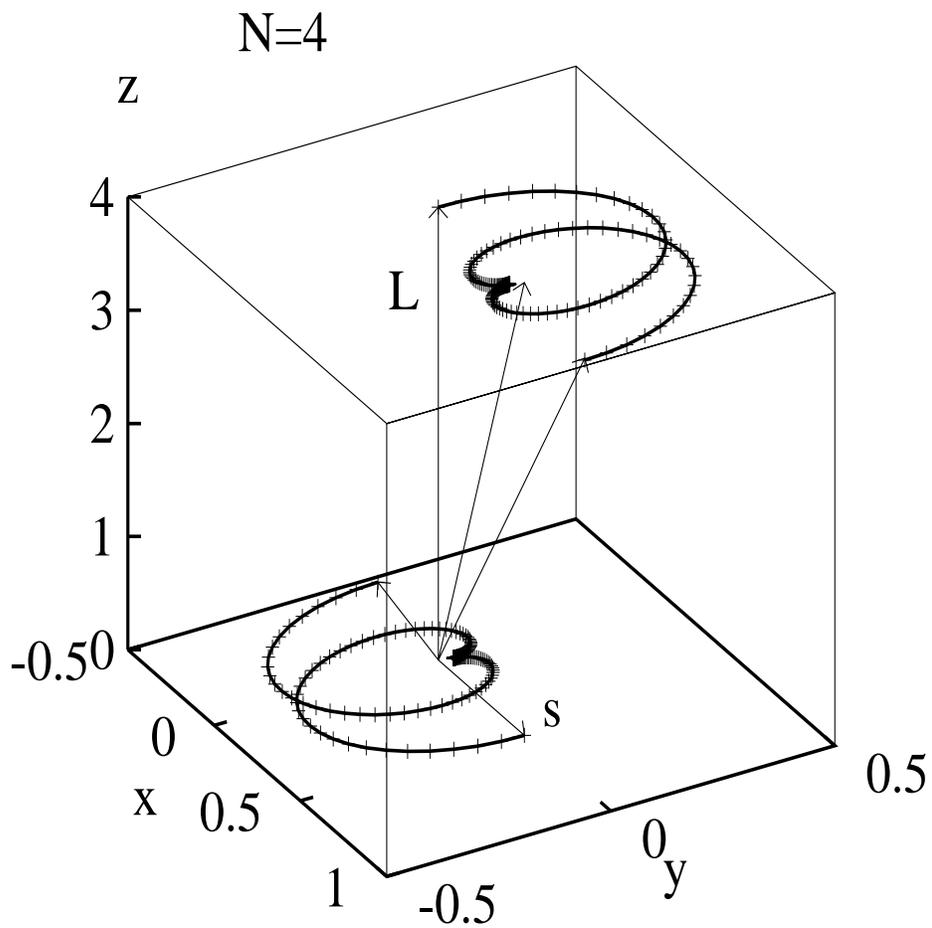

Fig.1. Motion of $\langle\vec{s}\rangle$ and $\langle\vec{l}\rangle$ during $T_{ls}/2$ illustrating oscilations and exchange of angular momenta (*spin-orbit pendulum*). Symbols show the equidistant time steps $T_{ls}/500$.

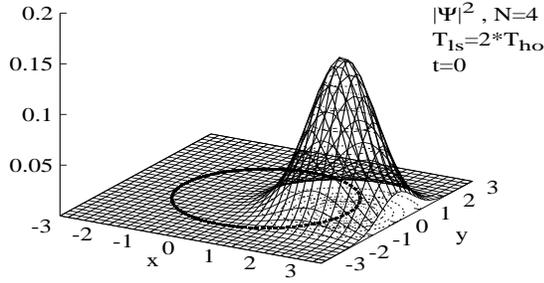

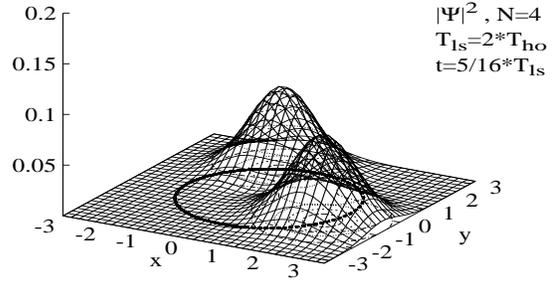

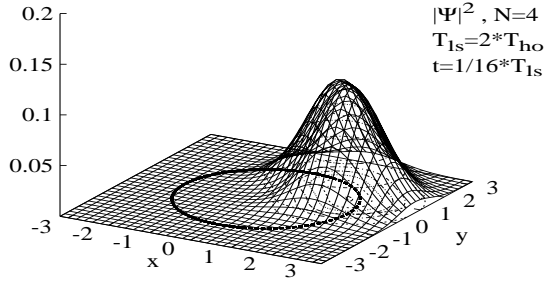

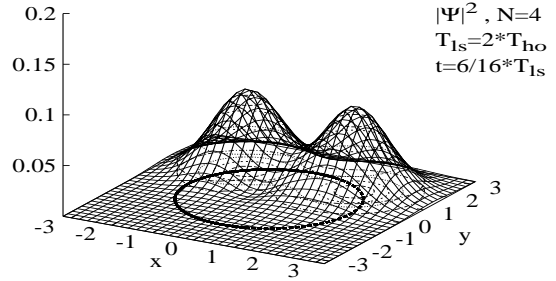

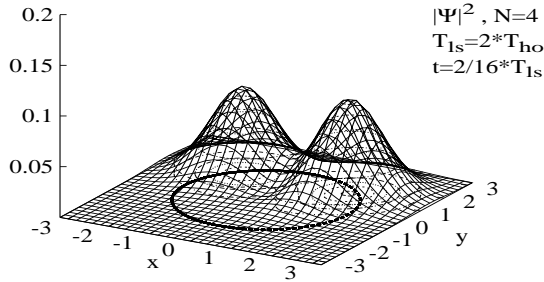

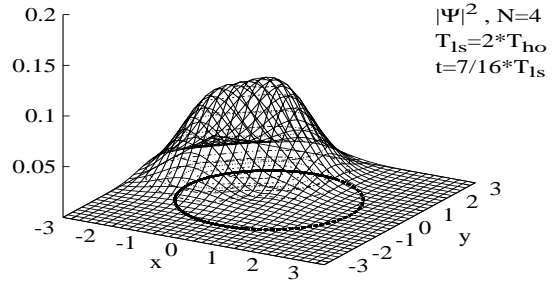

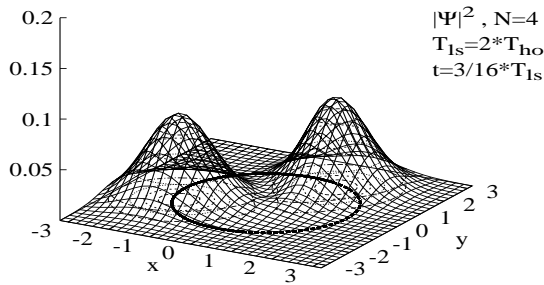

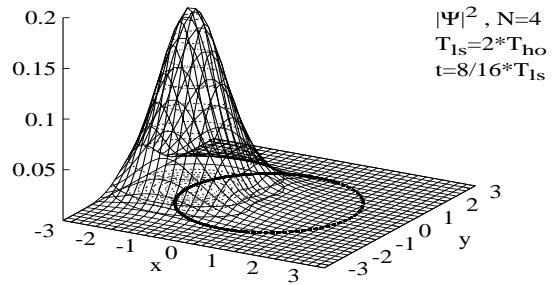

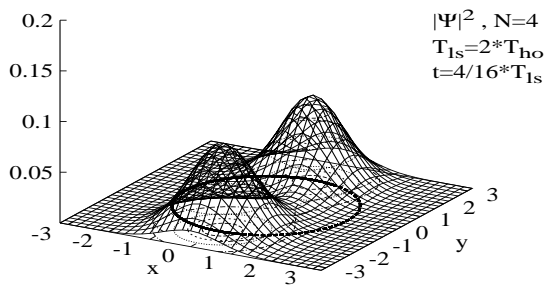

Fig. 3. Motion of the wave packet with $N=4$ in time range $[0, T_{ls}/2]$. Shown is $|\Psi(t)|^2 = |\Psi_+(t)|^2 + |\Psi_-(t)|^2$ integrated over $\theta$ as the function of coordinates on the plane of the classical orbit (marked by the thick circle) for $N = 4$. The time step is $\Delta t = \frac{1}{16} T_{ls}$. The strength of spin-orbit interaction is chosen to ensure $\omega_0 = 2\omega_{ls}$.

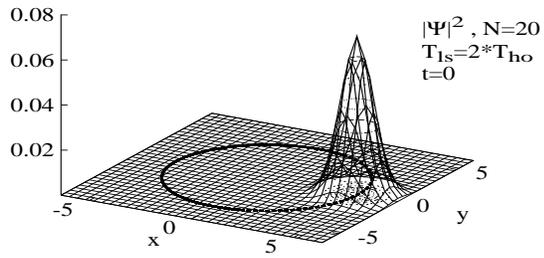

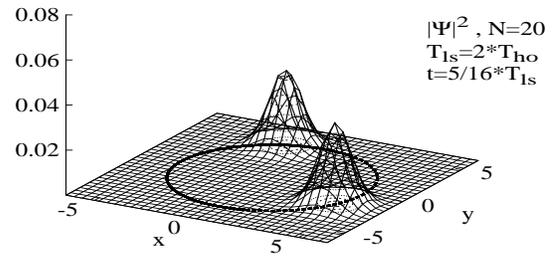

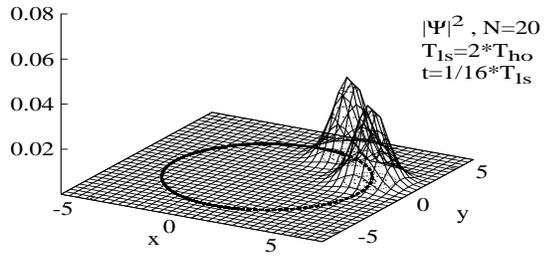

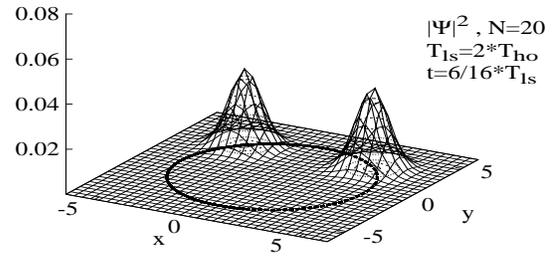

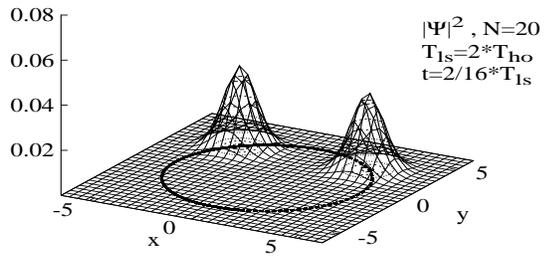

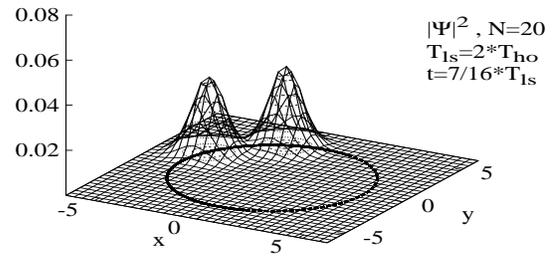

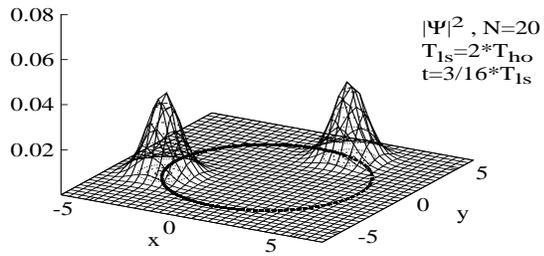

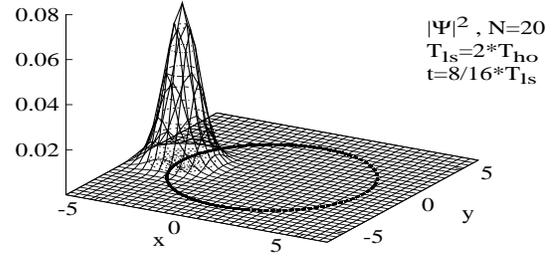

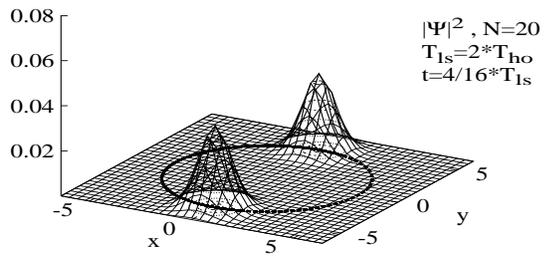

Fig. 3. The same as in Fig. 2 but for $N=20$ case.

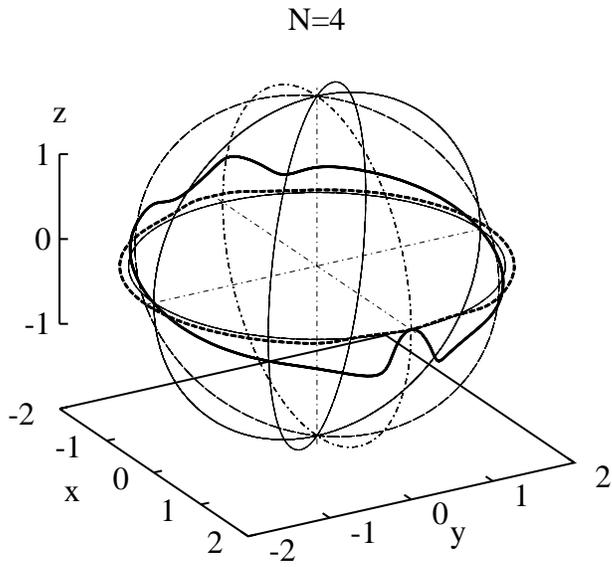

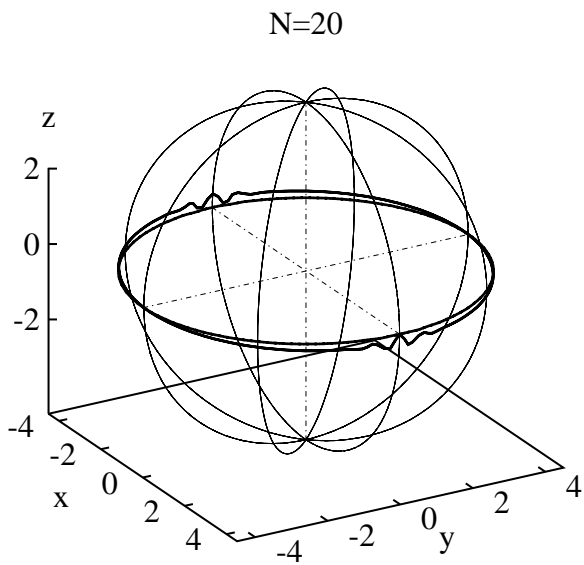

Fig. 4. Trajectories of the maximum of the subpacket with spin up (solid line) and that with spin down (dashed line) on the sphere (thin circles) with the radius equal to the radius of classical orbit. Top: case $N=4$, Bottom: case $N=20$.